\documentclass[twocolumn,english,aps,pra,reprint,showpacs]{revtex4-1}
\usepackage[T1]{fontenc}
\usepackage[latin9]{inputenc}
\usepackage{amsmath}
\usepackage{amssymb}
\usepackage{graphicx}
\usepackage{esint}

\makeatletter
\@ifundefined{textcolor}{}
{%
 \definecolor{BLACK}{gray}{0}
 \definecolor{WHITE}{gray}{1}
 \definecolor{RED}{rgb}{1,0,0}
 \definecolor{GREEN}{rgb}{0,1,0}
 \definecolor{BLUE}{rgb}{0,0,1}
 \definecolor{CYAN}{cmyk}{1,0,0,0}
 \definecolor{MAGENTA}{cmyk}{0,1,0,0}
 \definecolor{YELLOW}{cmyk}{0,0,1,0}
}

\usepackage{babel}

\makeatother

\usepackage{babel}
\begin{document}

\preprint{This line only printed with preprint option}

\title{Hydrodynamics of vortices in Bose-Einstein condensates:\\
 A defect-gauge field approach}

\author{F. Ednilson A. dos Santos}

\affiliation{Department of Physics, Federal University of São Carlos, 13565-905,
São Carlos, SP, Brazil}
\begin{abstract}
This work rectifies the hydrodynamic equations commonly used to describe
the superfluid velocity field in such a way that vortex dynamics
are also taken into account. In the field of quantum turbulence, it
is of fundamental importance to know the correct form of the equations
which play similar roles to the Navier-Stokes equation in classical
turbulence. Here, such equations are obtained by carefully taking
into account the frequently overlooked multivalued nature of the $U(1)$
phase field. Such an approach provides exact analytical explanations
to some numerically observed features involving the dynamics of quantum
vortices in Bose-Einstein condensates, such as the universal $t^{1/2}$
behavior of reconnecting vortex lines. It also expands these results
beyond the Gross-Pitaevskii theory so that some features can be generalized
to other systems such as superfluid $^{4}$He, dipolar condensates,
and mixtures of different superfluid systems. 
\end{abstract}

\pacs{67.85.De, 03.75.Lm, 67.25.dk}

\maketitle

\section*{Introduction}

Superfluidity is a macroscopic quantum phenomenon that has attracted
attention since its discovery in liquid $^{4}$He by Kapitza \cite{Kapitza1938}
as well as, Allen and Misener \cite{Allen1938}. London \cite{London1938}
proposed that the superfluidity appearing in $^{4}$He was closely
related to the existence of a Bose-Einstein condensate (BEC) which
can be described by a complex wavefunction $\psi=\sqrt{\rho}e^{iS}$,
where $\rho$ is the condensate density and $S$ is the phase that
determines the superfluid velocity, which is usually assumed to be
$\mathbf{v}\overset{?}{=}(\hbar/m){\boldsymbol\nabla} S$, where the notation $\overset{?}{=}$
represents the fact that this equation is not correct in general and
therefore must be modified, as we will see below. The multivalued
nature of $S$ implies the quantization of the superfluid vorticity
\cite{pethick2002bose,pitaevskii2016bose}. Since $S$ is defined
modulo $2\pi$, the velocity circulation $\ointclockwise\mathbf{v}\cdot d\mathbf{r}$
must be an integer multiple of $2\pi\hbar/m=h/m$.

Associations between quantum-vortex degrees of freedom and gauge fields
have been previously discussed in great detail by Kleinert \cite{kleinert1,kleinert2,kleinert3,kleinert4,kleinert2008multivalued},
where the concept of defect-gauge fields was introduced. An alternative
approach is discussed, for example, by Kozhevnikov \cite{PhysRevD.59.085003,Kozhevnikov2015122},
where vortex gauge fields are introduced as extra terms in the equations
of motion for the complex scalar field. An interesting possibility
is based on the exploration of approximate boson-vortex dualities
as in \cite{PhysRevD.14.1524,0295-5075-77-4-47005}, where the continuity
equation is used as a basis for the introduction of gauge fields.
In Ref. \cite{anglin}, a gauge field which is dual to the velocity
field is considered, thus allowing the study of the motion of a two-dimensional (2D)
point vortex in inhomogeneous backgrounds. In Ref. \cite{klein},
Popov's functional integral formalism \cite{popov}, where a gauge
field is introduced in order to enforce the constraint between velocity
and vorticity fields, is also applied to the study of  2D vortex motions. 

The analogy between quantum and classical hydrodynamics is usually
made by using the Gross-Pitaevskii (GP) equation

\begin{equation}
i\partial_{t}\psi=-\frac{1}{2}\nabla^{2}\psi+V(\mathbf{r})\psi+g\left|\psi\right|^{2}\psi,\label{eq:GP-1}
\end{equation}
where direct substitution of $\psi=\sqrt{\rho}e^{iS}$ seems to lead
to the hydrodynamic equations \cite{pethick2002bose,pitaevskii2016bose}:
\begin{eqnarray}
\partial_{t}\rho & = & -{\boldsymbol\nabla}\cdot(\rho\mathbf{v}),\label{eq:continuity}\\
\partial_{t}\mathbf{v} & \overset{?}{=} & {\boldsymbol\nabla}\left[\frac{1}{2}\left(\frac{1}{2\rho}\nabla^{2}\rho-\frac{1}{4\rho^{2}}\left|\nabla\rho\right|^{2}-\frac{v^{2}}{2}\right)-V-g\rho\right],\label{eq:hydro-false}
\end{eqnarray}
where for simplicity the system of units is chosen so that $\hbar=m=1$.
In Eq. (\ref{eq:hydro-false}), a usually unnoticed complication arises:
$S$ is a multivalued field and therefore the chain rule of differentiation
cannot be applied to $e^{iS}$ \cite{kleinert2008multivalued}. Indeed,
by taking the curl in Eq. (\ref{eq:hydro-false}), one would be left
with the false statement that vorticity has no dynamics, i.e., $\partial_{t}\boldsymbol{\omega}=\partial_{t}({\boldsymbol\nabla}\times\mathbf{v})=0$.
Thus, Eq. (\ref{eq:hydro-false}) turns out to be of little use for
dealing with situations where the dynamics of vorticity plays an important
role, as in the case of quantum turbulence \cite{Feynman1955,Henn2009,Skrbek2012,Barenghi2014b,Tsatsos20161}.
In the latter, it is common to interpret results through an analogy
between Eq. (\ref{eq:hydro-false}) and the Navier-Stokes equation,
thus establishing a close relationship between quantum and classical
turbulence \cite{Barenghi2014a,Tsatsos20161}. In practice, due to
the weaknesses of Eq. (\ref{eq:hydro-false}), studies are normally
based on direct numerical simulations of the GP equation, as in \cite{PhysRevLett.110.104501,Kobayashi2005,PhysRevA.93.033651},
or the Biot-Savart model, as in \cite{Hall1956a,Hall1956b,Schwarz1985}.

The present work aims to provide a general framework where exact hydrodynamic
equations can be obtained for models of superfluidity described by
complex fields which have equations of motion of the form

\begin{equation}
\frac{\partial\psi(\mathbf{r},t)}{\partial t}=\mathcal{F}\{\psi^{\ast},\psi\}(\mathbf{r},t),\label{eq:general}
\end{equation}
where $\mathcal{F}\{\psi^{\ast},\psi\}(\mathbf{r},t)$ can be any
arbitrary functional of $\psi$ which is local in time (i.e., depends
only on $\psi$ at the instant $t$) and has explicit dependency in
$\mathbf{r}$ and $t$. The Gross-Pitaevskii theory based on Eq. (\ref{eq:GP-1})
is therefore only one example. This is made possible through a careful
analysis of the multivalued nature of the phase field $S$, following
the lines presented in Ref. \cite{kleinert2008multivalued}. The hydrodynamic
equation obtained this way makes it possible to derivate the superfluid
behavior in its entirety, which includes its vorticity dynamics.

\subsection*{Two dimensional case}

For illustrative purposes, let us consider a scalar complex field
$\psi$ in two spatial dimensions with its usual definition for the
velocity field \cite{pethick2002bose,pitaevskii2016bose} 
\begin{equation}
\mathbf{v}=\frac{\psi^{\ast}{\boldsymbol\nabla}\psi-\psi{\boldsymbol\nabla}\psi^{\ast}}{2i\psi^{\ast}\psi}.
\end{equation}

\begin{figure*}
\begin{centering}
\includegraphics[scale=0.4]{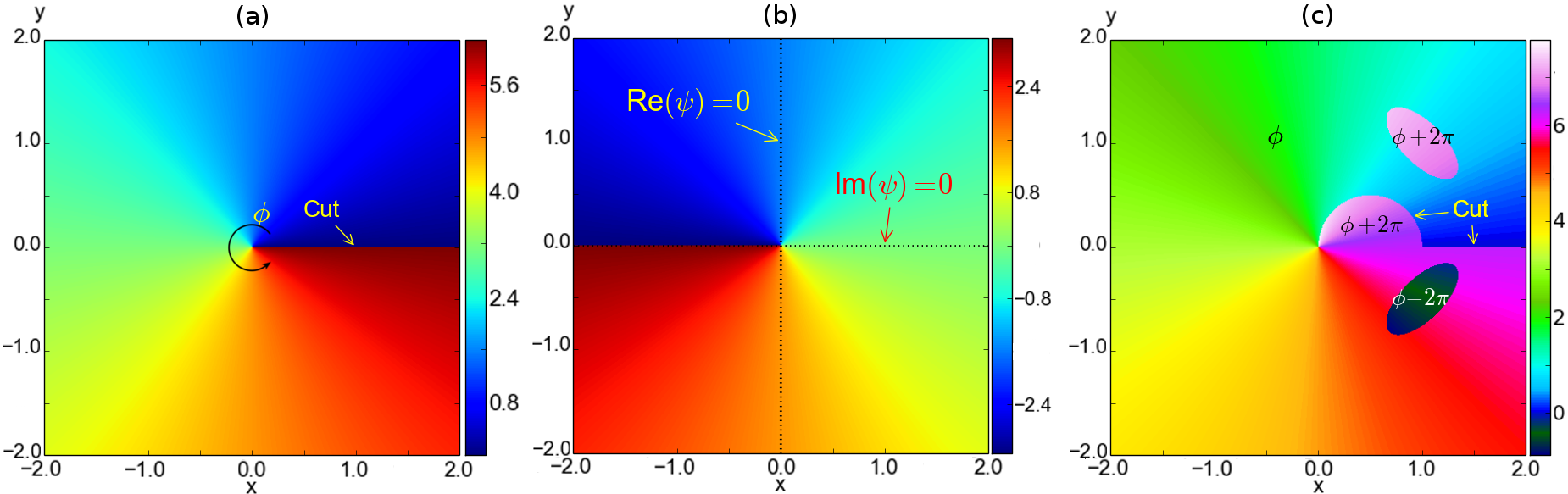} 
\par\end{centering}

\caption{\label{fig1} (color online) Different representations of the same multivalued phase
corresponding to the velocity field in Eq. (\ref{eq:v-2D}). In (a),
$S$ is defined so that $0\leq S<2\pi$. In (b), $S$ is defined so
that $-\pi\leq S<\pi$ and (c) corresponds to a gauge transformation
which adds $2\pi l$, with different $l\in\mathbb{Z}$ for different
regions of the plane.}
\end{figure*}

A common approach at this point would be to consider the Madelung's
representation $\psi=\sqrt{\rho}e^{iS}$ and use the chain rule of
derivatives in order to obtain the relation between the phase $S$
and velocity field $\mathbf{v}$, as follows: 
\begin{equation}
\mathbf{v}=\frac{\psi^{\ast}{\boldsymbol\nabla}\psi-\psi{\boldsymbol\nabla}\psi^{\ast}}{2i\psi^{\ast}\psi}=\frac{e^{-iS}{\boldsymbol\nabla} e^{iS}-e^{iS}{\boldsymbol\nabla} e^{-iS}}{2i}\overset{?}{=}{\boldsymbol\nabla} S.\label{eq:1}
\end{equation}
However, as pointed out by Kleinert in \cite{kleinert2008multivalued},
the chain rule should not be indiscriminately used in the case of
multivalued fields. A typical example is that of a 2D isotropic vortex
which, in polar coordinates, can be expressed as $\psi(\mathbf{r})=f(r)e^{i\varphi}$,
with $0\leq\varphi<2\pi$ as in Fig. \ref{fig1}(a). In that case,
the field $S=\varphi$ is discontinuous over the cut line {[}see Fig.
\ref{fig1}(a){]}, thus giving 
\begin{equation}
{\boldsymbol\nabla} S=\frac{\hat{\varphi}}{r}-2\pi\Theta(x)\delta(y)\hat{y},
\end{equation}
where $\Theta(x)$ and $\delta(y)$ are the Heaviside and Dirac functions,
respectively, while $\hat{\varphi}$ and $\hat{y}$ are the unit vectors
corresponding to $\varphi$ and $y$. Observe that, in this way, the
property ${\boldsymbol\nabla}\times{\boldsymbol\nabla} S=0$ is preserved as expected. However,
a direct calculation of the velocity field gives 
\begin{equation}
\mathbf{v}=\frac{\psi^{\ast}{\boldsymbol\nabla}\psi-\psi{\boldsymbol\nabla}\psi^{\ast}}{2i\psi^{\ast}\psi}=\frac{\hat{\varphi}}{r},\label{eq:v-2D}
\end{equation}
which means that 
\begin{eqnarray}
{\boldsymbol\nabla} S & = & \mathbf{v}-\mathbf{A},\\
\mathbf{A} & = & 2\pi\Theta(x)\delta(y)\hat{y}.
\end{eqnarray}
Therefore, formula (\ref{eq:1}) for the velocity field must be correctly
defined according to 
\begin{equation}
\mathbf{v}={\boldsymbol\nabla} S+\mathbf{A},
\end{equation}
where the vector field $\mathbf{A}$ compensates for the discontinuity
in $S$. In addition, all the vorticity of $\mathbf{v}$ is concentrated
in the field $\mathbf{A}$, i.e.,
\begin{equation}
{\boldsymbol\nabla}\times\mathbf{v}={\boldsymbol\nabla}\times\mathbf{A}=2\pi\delta(\mathbf{r})\hat{z}.
\end{equation}

In order to make these results consistent with Eq. (\ref{eq:1}),
the common chain rule of differentiation must be modified \cite{kleinert2008multivalued}
according to 
\begin{equation}
{\boldsymbol\nabla} e^{iS}=i\mathbf{v}e^{iS}=i\left({\boldsymbol\nabla} S+\mathbf{A}\right)e^{iS}.
\end{equation}

Due to the $U(1)$ symmetry of $\psi$, the definition of $S$ can
always be modified by adding to it a scalar field $Q$ which assumes
values equal to $2\pi l$, with $l\in\mathbb{Z}$, where $l$ can
be different for different regions of the plane {[}see Fig. \ref{fig1}(c){]}.
Observe also in Figs. \ref{fig1}(b) and \ref{fig1}(c) that cut lines can be moved
due to the extra $Q$ field. This way, the field $\mathbf{A}$ also
has to change in order not to modify $\mathbf{v}={\boldsymbol\nabla} S+\mathbf{A}$,
therefore $\mathbf{v}$ is invariant under the following gauge transformations
\begin{eqnarray}
S & \rightarrow & S+Q,\\
\mathbf{A} & \rightarrow & \mathbf{A}-{\boldsymbol\nabla} Q.
\end{eqnarray}

\subsection*{Arbitrary number of dimensions}

From now on, the tensor notation with Einstein summation rule will
be used, where Greek indices correspond to space-time coordinates
and Latin indices correspond to pure spatial coordinates.

The previous analysis can be extended to arbitrary space-time dimensions,
where we have the four-velocity field $v_{\mu}$ given by 
\begin{equation}
v_{\mu}=\frac{\psi^{\ast}\partial_{\mu}\psi-\psi\partial_{\mu}\psi^{\ast}}{2i\psi^{\ast}\psi}=\frac{e^{-iS}\partial_{\mu}e^{iS}-e^{iS}\partial_{\mu}e^{-iS}}{2i}=\partial_{\mu}S+A_{\mu},\label{eq:v-def}
\end{equation}
where $\partial_{\mu}e^{iS}=i(\partial_{\mu}S+A_{\mu})e^{iS}$ and
the gauge field $A_{\mu}$ must be chosen such that it accounts for
any artificial discontinuities from $\partial_{\mu}S$. This leads
to the gauge transformations 
\begin{eqnarray}
S & \rightarrow & S+Q,\label{eq:G1}\\
A_{\mu} & \rightarrow & A_{\mu}-\partial_{\mu}Q.\label{eq:G2}
\end{eqnarray}

\subsection*{Topological conservation laws}

In analogy to the electromagnetic theory, we can use the gauge field
$A_{\mu}$ to define the force field tensor 
\begin{eqnarray}
F_{\mu\nu} & = & \partial_{\mu}A_{\nu}-\partial_{\nu}A_{\mu}\label{eq:force1}\\
 & = & \partial_{\mu}v_{\nu}-\partial_{\nu}v_{\mu},\nonumber 
\end{eqnarray}
which is invariant under the gauge transformations (\ref{eq:G1})--(\ref{eq:G2}).
Such a definition leads to the conservation laws 
\begin{eqnarray}
\text{In (2+1) dimensions: }\partial_{\mu}\left(\frac{1}{2}\varepsilon^{\mu\alpha\beta}F_{\alpha\beta}\right) & = & 0,\label{eq:conservation2D}\\
\text{In (3+1) dimensions: }\partial_{\mu}\left(\frac{1}{2}\varepsilon^{\mu\nu\alpha\beta}F_{\alpha\beta}\right) & = & 0,\label{eq:conservation3D}
\end{eqnarray}
where $\varepsilon$ is the Levi-Civita symbol.

The topological charge density in (2+1) dimensions is the vorticity
\begin{equation}
\Omega^{0}=\frac{1}{2}\varepsilon^{0ij}F_{ij}=\varepsilon^{ij}\partial_{i}v_{j}=\omega,\label{eq:charge2D}
\end{equation}
while the vortex-current vector is 
\begin{equation}
\Omega^{i}=\frac{1}{2}\varepsilon^{i0j}F_{0j}+\frac{1}{2}\varepsilon^{ij0}F_{j0}=-\varepsilon^{ij}F_{0j}=-\varepsilon^{ij}E_{j}.\label{eq:current2D}
\end{equation}

In (3+1) dimensions the topological charge density is the vorticity
vector 
\begin{equation}
\Omega^{0i}=\frac{1}{2}\varepsilon^{0ijk}F_{jk}=\varepsilon^{ijk}\partial_{j}v_{k}=\omega^{i},
\end{equation}
while the vortex-current tensor is 
\begin{equation}
\Omega^{ij}=\frac{1}{2}\varepsilon^{ij0k}F_{0k}+\frac{1}{2}\varepsilon^{ijk0}F_{k0}=-\varepsilon^{ijk}F_{0k}=-\varepsilon^{ijk}E_{k}.
\end{equation}
Here, the field $E_{i}$ is the timelike component of the antisymmetric
force-field tensor which in the electromagnetic theory corresponds to the electric
field 
\begin{equation}
E_{i}\equiv F_{0i}\equiv\partial_{0}A_{i}-\partial_{i}A_{0}.
\end{equation}

\subsection*{Hydrodynamic equations}

Now, in order to derive the correct hydrodynamic equations, let us
consider the derivative 
\begin{equation}
\partial_{i}v_{0}=\partial_{i}\left(\partial_{0}S+A_{0}\right)=\partial_{0}\left(\partial_{i}S+A_{i}\right)+\partial_{i}A_{0}-\partial_{0}A_{i},
\end{equation}
which can be rearranged in order to give the time derivative of the
spacelike velocity field 
\begin{equation}
\partial_{t}v_{i}\equiv\partial_{0}v_{i}=E_{i}+\partial_{i}v_{0},\label{eq:hydro}
\end{equation}
where $v_{0}$ is given by (\ref{eq:v-def}). Observe that $\partial_{0}\psi$
in Eq. (\ref{eq:general}) can be used for the calculation of $v_{0}$
in terms of $\rho$ and $v_{i}$. In the case of the GP equation (\ref{eq:GP-1}),
we have 
\begin{equation}
v_{0}=\frac{1}{2}\left[\frac{1}{2\rho}\nabla^{2}\rho-\frac{1}{4\rho^{2}}\left|\nabla\rho\right|^{2}-\frac{v^{2}}{2}\right]-V-g\rho.
\end{equation}
This corrects the usual hydrodynamic equation (\ref{eq:hydro-false})
so that $E_{i}$ takes into account all possible vorticity effects.
Thus, the correct hydrodynamic equation following from GP Eq. (\ref{eq:GP-1})
differs from the classical Euler equation in two aspects: the quantum
pressure and the force $E_{i}$. The vorticity equations can be obtained
by taking the curl in Eq. (\ref{eq:hydro}), which ends up reproducing
the vorticity conservation laws already stated in Eqs. (\ref{eq:conservation2D})
and (\ref{eq:conservation3D}). Although it is necessary to correct
(\ref{eq:hydro-false}), a straightforward calculation shows that
the continuity equation (\ref{eq:continuity}) remains valid, despite
the discontinuities of $S$.

\subsection*{Explicit form of force fields}

In order to construct a full hydrodynamic theory, it is also necessary
to express the force fields in Eq. (\ref{eq:force1}) in terms of
$\rho$ and $v_{i}$. According to (\ref{eq:force1}), $F_{\mu\nu}$
can be obtained straightforwardly once an explicit form of the gauge
field $A_{\mu}$ is known. In order to do that, let us consider the
phase field $S$ as being restricted to $0\leq S<2\pi$, in analogy
to the two-dimensional example presented earlier. The discontinuities
appearing in $S$ must be compensated by the gauge field in order
to allow for the correct calculation of the velocity field, as defined
in Eq. (\ref{eq:v-def}). Considering that $R$ and $I$ are the real
and imaginary parts of $\psi$, respectively, such a convention for
$S$ implies that its discontinuities appear when $R\geq0$ and $I=0$.
This means that $\partial_{\mu}S$ will have discontinuities of the
form $-2\pi\Theta(R)\partial_{\mu}\Theta(I)$. Therefore, the gauge
field must be given by 
\begin{equation}
A_{\mu}=2\pi\Theta(R)\partial_{\mu}\Theta(I).\label{eq:gauge1}
\end{equation}
This leads directly to the force field 
\begin{eqnarray}
F_{\mu\nu} & = & 2\pi\delta(R)\delta(I)\left(\partial_{\mu}R\partial_{\nu}I-\partial_{\nu}R\partial_{\mu}I\right)\nonumber \\
 & = & i\pi\delta(R)\delta(I)\left(\partial_{\mu}\psi\partial_{\nu}\psi^{\ast}-\partial_{\nu}\psi\partial_{\mu}\psi^{\ast}\right).\label{eq:force2}
\end{eqnarray}
Finally, by using the property $\delta(R)\delta(I)=2\delta(R^{2}+I^{2})/\pi$,
we get the hydrodynamic form of the force field 
\begin{equation}
F_{\mu\nu}=2\delta(\rho)\left(\partial_{\mu}\rho v_{\nu}-\partial_{\nu}\rho v_{\mu}\right).\label{eq:force3}
\end{equation}
In such a way, the field $E_{i}=F_{0i}$, necessary in Eq. (\ref{eq:hydro}),
is 
\begin{equation}
E_{i}=-2\delta(\rho)\left[v_{i}\partial_{j}\left(\rho v_{j}\right)+v_{0}\partial_{i}\rho\right],
\end{equation}
where $\partial_{0}\rho$ is obtained from the continuity equation
Eq. (\ref{eq:continuity}) and $v_{0}$ is model specific.

\subsection*{Vortex motion}

As a testing ground to the validity of the theory presented, let us
check whether it is indeed capable of predicting the correct motion
of point vortices in 2D and vortex lines in three dimensions (3D). In fact, the motion
of vortex lines can be analyzed by looking at the motion of point
vortices over planes crossed by the vortex line. Hence this discussion
can be reduced to the two-dimensional situation.

In this case, the motion of vortices can be described by Eq. (\ref{eq:conservation2D}),
while vortex currents can be directly evaluated from (\ref{eq:force2}).
As illustrated in Fig. \ref{fig1}, a singly quantized vortex is always
located at the crossing between $R=0$ and $I=0$ lines in the $xy$
plane. Let us consider, without loss of generality, that such a crossing
happens at the origin. At the vicinity of the crossing point, the
$\delta$ functions in (\ref{eq:force2}) can then be simplified to 
\begin{equation}
\delta(R)\delta(I)=\frac{\delta(x)\delta(y)}{\left|\varepsilon^{ij}\partial_{i}R\partial_{j}I\right|}.\label{eq:deltas1}
\end{equation}
From (\ref{eq:charge2D}), (\ref{eq:force3}), and (\ref{eq:deltas1})
we get 
\begin{eqnarray}
\omega=\Omega^{0} & = & 2\pi\delta(x)\delta(y)\frac{\epsilon^{ij}\partial_{i}R\partial_{j}I}{\left|\varepsilon^{lm}\partial_{l}R\partial_{m}I\right|}\nonumber \\
 & = & 2\pi\,{\rm sgn}\left(\epsilon^{ij}\partial_{i}R\partial_{j}I\right)\delta(x)\delta(y),
\end{eqnarray}
where ${\rm sgn}\left(\epsilon^{ij}\partial_{i}R\partial_{j}I\right)$
gives the vortex sign. Now combining (\ref{eq:current2D}), (\ref{eq:force3}),
and (\ref{eq:deltas1}), we have the vortex current 
\begin{equation}
\Omega^{i}=-2\pi\delta(x)\delta(y)\epsilon^{ij}\frac{\partial_{0}R\partial_{j}I-\partial_{j}R\partial_{0}I}{\left|\varepsilon^{lm}\partial_{l}R\partial_{m}I\right|}=\omega w^{i},
\end{equation}
where the vortex velocity $w^{i}$ is 
\begin{equation}
w^{i}=-\varepsilon^{ij}\frac{\partial_{0}R\partial_{j}I-\partial_{j}R\partial_{0}I}{\varepsilon^{lm}\partial_{l}R\partial_{m}I}.\label{eq:w-vortex}
\end{equation}
Observe that $w^{i}$ is indeed consistent with the equations describing
the motion of the crossing point between the lines $R=0$ and $I=0$, 
as in Ref. \cite{PhysRevA.61.032110}: 
\begin{eqnarray}
\partial_{0}R+w^{i}\partial_{i}R & = & 0,\\
\partial_{0}I+w^{i}\partial_{i}I & = & 0,
\end{eqnarray}
whose solution for $w_{i}$ is given by (\ref{eq:w-vortex}).

An elegant approximation for $w_{i}$ can be obtained for the case
of quasi-isotropic vortices with dynamics given by the GP equation
(\ref{eq:GP-1}), i.e., when 
\begin{equation}
\psi\approx\left[(x-x_{0})+i(y-y_{0})\right]\phi,\label{eq:quasi-iso}
\end{equation}
with $\phi=Ae^{i\lambda}$, where both $A$ and $\lambda$ have their
values as well as their first derivatives well defined at $(x,\, y)=(x_{0},\: y_{0})$.
By directly substituting (\ref{eq:quasi-iso}) into (\ref{eq:w-vortex})
and observing that at the vortex location we have $\partial_{0}\psi=i\frac{1}{2}\nabla^{2}\psi$,
we get 
\begin{equation}
w^{i}=\partial^{i}\lambda+\varepsilon^{ij}\partial_{j}\ln(A).\label{eq:v-motion}
\end{equation}
This gives a correction to the so-called point-vortex model, where
$\partial^{i}\lambda$ is the velocity field over the vortex core
excluding the self-generated velocity field, while $\varepsilon^{ij}\partial_{j}\ln(A)$
gives a contribution perpendicular to the density gradient. The necessity
for this correction has already been observed in the numerical studies
of Ref. \cite{PhysRevA.77.032107}.

Our 2D analysis can be directly generalized to 3D vortex lines by
considering a plane crossed by the vortex line. In this case, $w^{i}$
would describe the motion of the crossing point over the considered
plane.

\begin{figure*}
\begin{centering}
\includegraphics[scale=0.39]{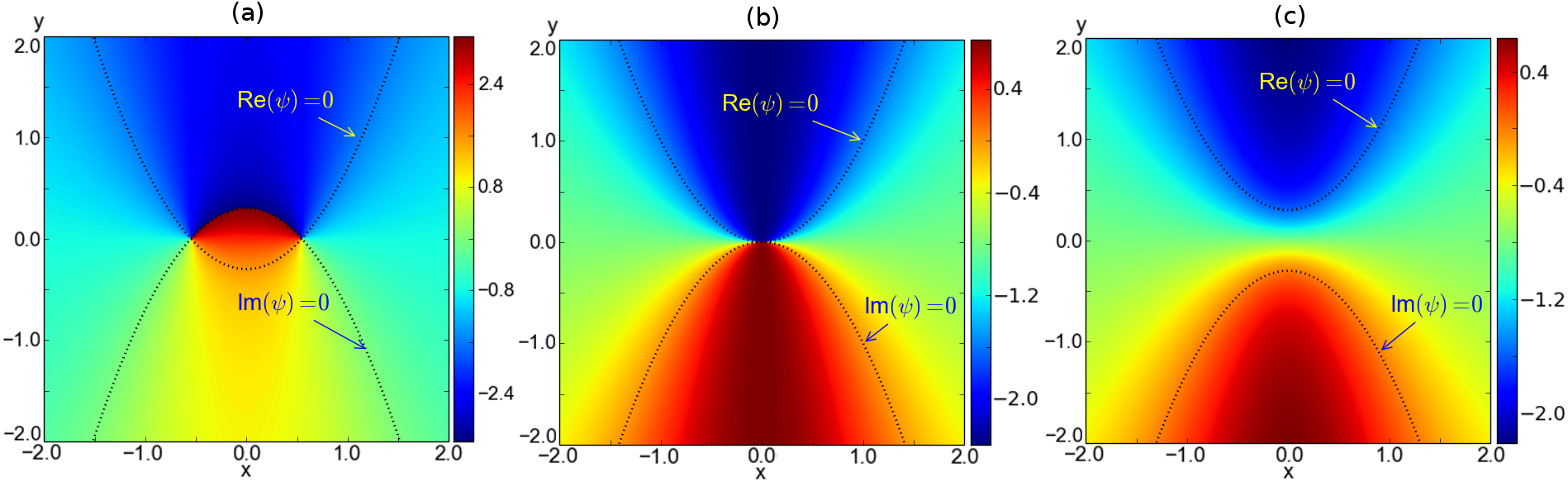} 
\par\end{centering}

\caption{\label{fig2} (color online) Annihilation or creation of a vortex pair. The sequence
(a)-(b)-(c) corresponds to the annihilation process, while the inverse
sequence (c)-(b)-(a) corresponds to the pair-creation process. Such sequences
also describe the recombination process of 3D vortex lines, where
the depicted plane crosses the recombination point.}
\end{figure*}

\subsection*{Reconnection of Lines and Creation/Annihilation of pairs}

An interesting situation occurs when the $R=0$ and $I=0$ lines touch
each other tangentially at a single point as in Fig. \ref{fig2}(b).
Actually, Fig. \ref{fig2} can illustrate either the situation right
at the beginning of a vortex-pair creation process or at the end of
a vortex-pair annihilation process. Indeed, the sequence (a)-(b)-(c)
in Fig. \ref{fig2} exemplifies a vortex-pair annihilation process,
while the inverse sequence (c)-(b)-(a) describes a pair-creation process.
For simplicity, without loss of generality, one can consider that
the lines touch at $x_{0}=y_{0}=t_{0}=0$ and are tangent to the $x$ axis
at this point, i.e., $\partial_{x_{0}}R=\partial_{x_{0}}I=0$. At the
vicinity of the touching point the Taylor expansion can be used: 
\begin{eqnarray}
R & = & \frac{\partial R}{\partial y_{0}}y+\frac{\partial R}{\partial t_{0}}t+\frac{1}{2}\frac{\partial^{2}R}{\partial x_{0}^{2}}x^{2}+\cdots,\label{eq:c1}\\
I & = & \frac{\partial I}{\partial y_{0}}y+\frac{\partial I}{\partial t_{0}}t+\frac{1}{2}\frac{\partial^{2}I}{\partial x_{0}^{2}}x^{2}+\cdots.\label{eq:c2}
\end{eqnarray}
Close to the touching point, the curves $R=0$ and $I=0$ can then
be obtained by considering the dominant terms in (\ref{eq:c1}) and
(\ref{eq:c2}), according to 
\begin{eqnarray}
y_{{\rm Re}} & \approx & -\frac{1}{2}\frac{\partial^{2}R/\partial x_{0}^{2}}{\partial R/\partial y_{0}}x^{2}-\frac{\partial R/\partial t_{0}}{\partial R/\partial y_{0}}t,\\
y_{{\rm Im}} & \approx & -\frac{1}{2}\frac{\partial^{2}I/\partial x_{0}^{2}}{\partial I/\partial y_{0}}x^{2}-\frac{\partial I/\partial t_{0}}{\partial I/\partial y_{0}}t.
\end{eqnarray}
The crossing points as depicted in Fig. \ref{fig2}(a) are solutions
of the condition $y_{{\rm Re}}=y_{{\rm Im}}$, which are given by
\begin{eqnarray}
x^{2} & \approx & 2t\alpha,\label{eq:m-pair}\\
\alpha & = & \left.\frac{\partial_{t}R\partial_{y}I-\partial_{t}I\partial_{y}R}{\partial_{x}^{2}I\partial_{y}R-\partial_{x}^{2}R\partial_{y}I}\right|_{x=y=t=0}.
\end{eqnarray}
The sign of $\alpha$ indicates whether there are real solutions for
$x$ with $t<0$ or with $t>0$, thus determining if it is the case
of an annihilation ($\alpha<0$) or creation ($\alpha>0$) process.
Also from (\ref{eq:m-pair}), we get the power-law behavior for the
creation or annihilation process:
\begin{equation}
x\sim\pm t^{1/2}.
\end{equation}
Observe that these results can also be directly obtained from (\ref{eq:w-vortex})
by considering the expansions (\ref{eq:c1}) and (\ref{eq:c2}) and
neglecting the subdominant terms. This calculation would then lead
to 
\begin{eqnarray}
w^{y} & \approx & -\left.\frac{\partial_{t}R\partial_{x}^{2}I-\partial_{t}I\partial_{x}^{2}R}{\partial_{x}^{2}I\partial_{y}R-\partial_{x}^{2}R\partial_{y}I}\right|_{x=y=t=0},\\
w^{x} & \approx & \frac{1}{x}\left(\left.\frac{\partial_{t}R\partial_{y}I-\partial_{t}I\partial_{y}R}{\partial_{x}^{2}I\partial_{y}R-\partial_{x}^{2}R\partial_{y}I}\right|_{x=y=t=0}\right)=\frac{\alpha}{x}.
\end{eqnarray}

Creation and annihilation of vortex pairs also leave their signatures
in the hydrodynamic equation (\ref{eq:hydro}). A direct evaluation
of $F_{\mu\nu}$ at $t=0$ and around the point $x_{0}=y_{0}=0$ can
be obtained with the help of Eqs. (\ref{eq:c1}) and (\ref{eq:c2}). It
then gives 
\begin{equation}
F_{\mu\nu}=4\pi\delta(x^{2})\delta(y)\frac{\partial_{\mu}R\partial_{\nu}I-\partial_{\nu}R\partial_{\mu}I}{\left|\partial_{x}^{2}R\partial_{y}I-\partial_{x}^{2}I\partial_{y}R\right|}.
\end{equation}
Since $\partial_{x}R=\partial_{x}I=0$ at $x=y=0$, the vorticity
$\omega=\varepsilon^{ij}F_{ij}$ vanishes. However, it does not mean
that the vorticity flux vanishes in all directions. Actually, the
vorticity flux in the $y$-direction $\Omega^{2}=F_{01}$ vanishes,
while for the $x$-direction we have 
\begin{equation}
\Omega^{1}=-F_{02}=-4\pi\delta(x^{2})\delta(y)\frac{\partial_{t}R\partial_{y}I-\partial_{y}R\partial_{t}I}{\left|\partial_{x}^{2}R\partial_{y}I-\partial_{x}^{2}I\partial_{y}R\right|}.\label{eq:flux-x}
\end{equation}
This reflects the fact that although no vortex actually exists at
$t=0$, a vorticity flux is still necessary to account for the creation
and annihilation of vortex pairs occurring in the superfluid. Also
the possibility of having a nonzero $E_{i}=F_{0i}$, even in the
absence of vortices, shows that the hydrodynamic equation (\ref{eq:hydro})
is capable of describing the creation and annihilation of vortex pairs.

Again, it should be emphasized that such a two-dimensional analysis
can also be directly generalized to the case of recombinations of
3D vortex lines by considering planes crossed by the vortex lines.
Indeed, the present analysis demonstrates exactly the $x\sim\pm t^{1/2}$
behavior for the reconnection of vortex lines which was observed experimentally
in Ref. \cite{Paoletti20101367}, numerically in the context of Biot-Savart
models in Ref. \cite{PhysRevLett.55.1749}, and analytically in the
context of GP equation in Ref. \cite{Nazarenko2003}. In addition,
such a $t^{1/2}$ law turns out to be very general in the sense that
it is not restricted to any particular superfluid model such as the
GP equation. Indeed, this result depends only on the existence of
the first time derivative as well as the first and second spacial
derivatives of $\psi$.

\subsection*{Conclusions}

This work provides a general framework for the construction of hydrodynamic
theories which are capable of correctly including any possible vortex
dynamics that may exist in a large set of superfluidity models. By
a detailed examination of the role of the multivalued nature of the
phase field $S$ in the vortex dynamics, the general hydrodynamic
equation (\ref{eq:hydro}) was obtained, where all details of a specific
model are introduced through the quantity $v_{0}$, defined in Eq.
(\ref{eq:v-def}). Such multivaluedness of $S$ demands the introduction
of the gauge field $A_{\mu}$, where the time-like component $E_{i}=F_{0i}$
of its force field must be introduced in the hydrodynamic equation
(\ref{eq:hydro}). The only restriction of this approach is that the
equation of motion for the macroscopic wave function $\psi$ must
be of first order in time, according to Eq. (\ref{eq:general}). As
a test for the practicality of this approach, the dynamics of 2D point
vortices and 3D vortex lines have been considered. It turns out that
the numerically observed behavior \cite{PhysRevA.77.032107} of point
vortices moving over a background density gradient is analytically
reproduced in Eq. (\ref{eq:v-motion}). In addition, the $t^{1/2}$
behavior of creation or annihilation of 2D vortex pairs as well as of
3D vortex line reconnections \cite{Paoletti20101367,PhysRevLett.55.1749,Nazarenko2003}
is exactly demonstrated for a large class of superfluid models.

\subsection*{Acknowledgments}

Acknowledgements to the National Council for the improvement of Higher
Education (CAPES). I wish also to thank A. Novikov, M. C. Tsatsos,
and Axel Pelster for reading and commenting on this manuscript.  \bibliographystyle{unsrt}
\bibliography{Gauge}

\end{document}